# The Star Formation Camera

## Paul Scowen

School of Earth & Space Exploration
Arizona State University
PO Box 871404, Tempe, AZ 85287-1404

(480) 965-0938

paul.scowen@asu.edu

Rolf Jansen (Arizona State University)
Matthew Beasley (University of Colorado – Boulder)
Daniela Calzetti (University of Massachusetts)
Steven Desch (Arizona State University)
Alex Fullerton (STScI)
John Gallagher (University of Wisconsin – Madison)
Doug Lisman (Jet Propulsion Laboratory, Caltech)
Steve Macenka (Jet Propulsion Laboratory, Caltech)
Sangeeta Malhotra (Arizona State University)
Mark McCaughrean (University of Exeter)
Shouleh Nikzad (Jet Propulsion Laboratory, Caltech)
Robert O'Connell (University of Virginia)
Sally Oey (University of Michigan)
Deborah Padgett (IPAC / Caltech)
James Rhoads (Arizona State University)
Aki Roberge (NASA – GSFC)
Oswald Siegmund (SSL – UC Berkeley)
Stuart Shaklan (Jet Propulsion Laboratory, Caltech)
Nathan Smith (University of California – Berkeley)
Daniel Stern (Jet Propulsion Laboratory, Caltech)
Jason Tumlinson (STScI)
Rogier Windhorst (Arizona State University)
Robert Woodruff (LMCO)

Activity White Paper for the Astro2010 Decadal Survey
Subcommittee on Programs

**Abstract**

The Star Formation Camera (SFC) is a wide-field (~15'×19', >280 arcmin$^2$), high-resolution (18×18 mas pixels) UV/optical dichroic camera designed for the *Theia* 4-m space-borne space telescope concept. SFC will deliver diffraction-limited images at λ > 300 nm in both a blue (190-517nm) and a red (517-1075nm) channel simultaneously. Our aim is to conduct a comprehensive and systematic study of the astrophysical processes and environments relevant for the births and life cycles of stars and their planetary systems, and to investigate and understand the range of environments, feedback mechanisms, and other factors that most affect the outcome of the star and planet formation process. Via a 4-Tier program, we will step out from the nearest star-forming regions within our Galaxy (Tier 1), via the Magellanic Clouds and Local Group galaxies (Tier 2), to other nearby galaxies out to the Virgo Cluster (Tier 3), and on to the early cosmic epochs of galaxy assembly (Tier 4). Each step will build on the detailed knowledge gained at the previous one. This program addresses the origins and evolution of stars, galaxies, and cosmic structure and has direct relevance for the formation and survival of planetary systems like our Solar System and planets like Earth. We present the design and performance specifications resulting from the implementation study of the camera, conducted under NASA's Astrophysics Strategic Mission Concept Studies program, which is intended to assemble realistic options for mission development over the next decade. The result is an extraordinarily capable instrument that will provide deep, high-resolution imaging across a very wide field enabling a great variety of community science as well as completing the core survey science that drives the design of the camera. The technology associated with the camera is next generation but still relatively high TRL, allowing a low-risk solution with moderate technology development investment over the next 10 years. We estimate the cost of the instrument to be $390M FY08.

**Acknowledgment**: A portion of the work described in this report was performed at the Jet Propulsion Laboratory, California Institute of Technology, under a contract with the National Aeronautics and Space Administration. © 2009. All rights reserved. Work was also performed at Arizona State University and Goddard Space Flight Center under contract NNX08AK79G from the National Aeronautics and Space Administration under the NRA entitled "Astrophysics Strategic Mission Concept Studies".

**Key Science Goals**

***Tier 1 – cf. Scowen et al SWP, "Understanding Global Galactic Star Formation"***

We believe that to understand and address star formation as a global *system*, we need to design and engage in a systematic program of imaging that covers a large number and variety of Galactic star forming regions. To understand star birth in the early Universe, to understand galaxy formation and evolution, to understand the origin of the stellar mass spectrum, to understand the formation of planets, and to understand feedback, we must treat star birth as an integrated systemic process. We must observe star forming complexes in their entirety: we must trace the interactions between gas and stars, between stars and stars, and between disks and their environments. To make progress, we must spatially resolve disks, multiple stars, and star clusters. We must measure stellar motions, and perform relative photometry with sufficient precision to age-date young stars. All these top-level goals make specific requirements of any instrumentation designed to execute this program – requirements that we will detail in subsequent sections. At the heart of this program is the goal of providing critical advances in our knowledge of star and planet birth.

The goals of our Galactic star forming imaging program are to make major advances in the following topics:

**Young stellar objects** (YSOs): *Masses, mass-spectra, rotation rates, variability, ages, multiplicity, clustering statistics, motions, brown dwarfs, free-floating proto-planets*. We need to be able to trace individual star, multiple star, and cluster properties to assay the range of star formation products and the manner in which they are assembled. Of particular interest is the search for transiting proto-planets in a subset of edge-on disks, and with a large-area imaging survey we will capture extremely rare types of events such as proto-planet collisions in 1 to 100 Myr old debris disks in associations.

**Disks**: *Sizes, masses, structure, mass-loss rates, photo-evaporation, density distributions, survival times*. A primary goal is to identify thousands of protoplanetary disks seen in silhouette, and embedded within evaporating proplyd envelopes in dozens of nearby HII regions, out to a distance of about 2 kpc. The widefield survey images taken toward regions such as Orion or Carina will extend the surveyed areas by one to two orders of magnitude over the most ambitious HST surveys undertaken so far.

**Outflows**: *Microjets, jets, wide-angle flows, winds, motions, momenta, mass-loss rates, turbulence, shocks*. HST has demonstrated that sub-arcsecond imaging is needed to begin to resolve the structure of shocks, and distinguish shock fronts from post-shock cooling layers. Furthermore, only space-based UV/optical observations can measure proper motions on a time-scale short compared to the cooling time. The survey observations will measure the proper motions of hundreds of outflows, enabling the first direct measure of the momentum and energy injected into the ISM by protostellar outflows for a wide-range of stellar masses and star forming environments.

**Nebulae**: *Excitation, motion, ionization fronts, triggered star formation*. This imaging program is deliberately designed to investigate the formation of HII regions and expanding bubble systems. How do ionization fronts disrupt surrounding clouds? Under what conditions to they trigger star formation in the medium? High spatial-resolution images with multiple narrowband filters are essential to resolve and correctly model the complex stratified structure of an ionization front.

**Massive stars**: *Motions, variations, winds, interactions with siblings, HII regions*. The program will also investigate stellar wind bubbles in HII regions and the interactions of stellar winds with cometary clouds, proplyds, naked young stars and their winds and jets, and the surrounding ISM.

**Recycling**: *Supernova remnants and planetary nebulae, bulk motions, excitation, shocks*. The late stages of stellar evolution - especially in massive stars - are an integral piece of the star and planet formation puzzle, because outflows from the deaths of massive stars drive the chemical evolution and energetics of the ISM. In particular, supernova ejecta enrich the ISM with the elements needed for life to exist, while supernova shocks and stellar winds may compress the surrounding ISM to trigger new star formation. Outflows from the deaths of intermediate mass stars (planetary nebulae) also enrich the ISM with dust, which is vital to the formation of molecular clouds.

**Superbubbles**: *Destruction of clouds, OB associations, T associations, global structure and evolution of star forming regions*. The energy input from the combined influence of UV radiation, stellar winds, and supernovae from massive stars makes "swiss cheese" out of the ISM. In the most massive star forming regions, where dozens of OB stars live fast and die young before moving very far from their birth sites, the combined effect of this feedback can blow giant shells or "superbubbles" that may eventually break out of the galactic plane, driving a galactic fountain that is vital to the recycling of the ISM.

**The Galactic Ecology**: *Impact of spiral arms, formation of clouds, Galactic gradients in YSO and cluster properties, the Galactic Center*. We believe an investigation of the "galactic ecology" is vital to understanding the global nature of the star formation process - the formation of giant molecular clouds from the ISM. How do HII regions and superbubble ionization fronts compress the surrounding ISM? Does ram-pressure trigger cloud formation? How do spiral arms trigger cloud formation? How do clouds and cloud cores collapse into clusters, and multiple stars?

***Tier 2 – cf. Scowen et al SWP, "The Magellanic Clouds Survey – a Bridge to Nearby Galaxies"***

***Feedback from Massive Stars*** Massive OB stars have a profound influence on their environment, ranging from destructive evaporation of molecular clouds that curtails further star formation, to galactic-scale production of ionizing radiation, galactic superwinds, and heavy elements that drive evolutionary processes in galaxies and the cosmos itself. The Magellanic Clouds, owing to their proximity and minimal Galactic obscuration, are a superior test-bed in which to examine both triggering and feedback processes on both microscopic and macroscopic scales.

With such a deep, narrowband imaging survey of the Clouds, the elusive large-scale ionization structure of the diffuse, warm ionized medium (DWIM) will become dramatically more apparent, allowing its spatial and ionization properties to be readily correlated with embedded star-forming regions, which presently are presumed to be the origin of the DWIM. This survey will also offer important leverage on the DWIM properties with respect to 3-D ISM structure and metallicity between the LMC and SMC. Comparison with the quantitative field massive star populations, as well as those in OB associations (see below) will provide unprecedented constraints on feedback parameters such as ionizing fluxes, stellar wind power, and elemental enrichment. It will be possible, with this survey dataset, to quantify and parameterize the spheres of influence of massive stars as a function of mass and interstellar conditions, for the three feedback effects.

***30 Doradus: The Nearest Giant Extragalactic HII Region*** The 30 Doradus Nebula plays a key role in our understanding of HII regions. In nearby regions within our own Galaxy, we can study the physical processes in detail. Work on M16 has shown that emission within the nebula arises predominantly within a narrow region at the interface between the HII region and the molecular cloud. However, an HII region like M16 is tiny in comparison with giant HII regions, and no giant HII regions are close enough to allow the stratified ionization structure of the photoevaporative flow to be studied directly. 30 Doradus alone offers an opportunity to bootstrap the physical understanding of small nearby HII regions into the context of the giant regions seen in distant galaxies. 30 Doradus is in a very different class of object from objects such as the Orion Nebula in terms of structure, size, dynamics, level of star formation, diversity of morphologies, and so on - we cannot learn what we need to know about Giant Extragalactic HII Regions from the study of simple Galactic HII regions. It is critical that we learn about 30 Doradus in and of itself.

***The Clustering of Star Formation*** Because of obscuration by dust, we have no clear view and understanding of star formation on large and global scales within our Galaxy. Such understanding must come from the study of other galaxies. On large scales, perhaps the most basic concept is that of a coherence length. If star formation proceeds in pockets that are independent of one another, then there should be no correlation of ages and distances between the different regions. Alternatively, if star formation propagates in a 'wave' from one side of the galaxy to the other, then there should be a linear correlation. When a star formation 'wave' dies out on scales larger than this distances (the coherence length), then the correlation between average age and separation disappears. Ground-based studies indicate that within the LMC ages and separations between star clusters are strongly correlated up to separations of ~1°. The correlation vanishes at larger separations, perhaps because the coherence length is limited by the thickness of the LMC disk or by the Jeans length.

***Tier 3 – cf. Jansen et al SWP, "A Systematic Study of the Stellar Populations and ISM in Galaxies out to the Virgo Cluster"***

***Near Field Cosmology: the oldest stellar populations***. It is useful to ask where the oldest stars are located. We know that in the Milky Way they reside in the spheroidal halo, in the LMC and SMC they have the largest radial scale of any stellar population, and they usually are the least centrally concentrated stellar component of dwarf spheroidals. However, even in nearby galaxies these results only apply in a mean sense. With the growing realization of the importance of interactions in the lives of galaxies, as demonstrated by the discovery of tidal debris streams and plumes in, e.g., the Milky Way, M31, M81, and NGC4013, the old star distribution merits reexamination. Are older stars asymmetrically distributed in the outer regions of galaxies, as expected if they were contributed by dissolving satellites?

***Star Formation and its Products***. The existing combination of ground-based and HST imaging provides an excellent base from which to design more ambitious investigations of the nature and extent of star forming sites. Investigations of connections between drivers, if any, for star formation — spiral arms, interactions, etc., as well as basic galactic properties — are essential for understanding how feedback operates. A survey of the local D<20Mpc volume provides the range of galaxy types, luminosities, cosmic environment, and the sensitivity and statistics to support a complete study of the association of compact clusters and regions of star formation. By combining deep mid-UV and narrow-band Hα observations, it becomes possible to also address the escape fraction of ionizing radiation in a variety of galaxies.

***Are Galactic Disks Growing?*** As already demonstrated by GALEX, young stars have high contrast against the sky in the mid-UV. This spectral range therefore opens the way for mapping star formation in low-density environments, including the outer disks of galaxies. A next-generation wide-field UV–near-IR space observatory must offer major advantages in sensitivity and resolution over the pioneering results from GALEX. Hence, we would be able to determine ages and photometric stellar masses for small star forming complexes of the type that appear to populate the outer disks of galaxies, ranging from small irregulars to giant spirals. From these, star formation rates per unit area and, thus, disk growth rates can be estimated.

***Galactic Centers***. Centers of galaxies are dumping grounds. Baryonic material that ends up in the central zone of a galaxy has experienced substantial dissipation and loss of angular momentum. We would be able to systematically chart the stellar properties of nuclear environments. Where and in what ways are stars formed (clusters, scaled OB associations, spiral arms, rings, clumps)? How does star formation relate to the properties of nuclei on small scales and on the other side to the surrounding main disk? How are bars, both large and nuclear, related to the structure and activity levels in nuclei?

***A Survey of Nearby Galaxies***. We propose to learn how galaxies work, through studies of their stars, ISM, and immediate environments, and to build the definitive UV–near-IR photometric imaging database of galaxies within our local slice of the Universe. This would result in a 21st century digital 'Hubble Atlas' of nearby galaxies and their surroundings that will provide a standard for testing our understanding of how galaxies attained their present forms and how their stellar components will likely evolve into the future. The resolved and unresolved stellar populations would be analyzed through color-magnitude and color-color diagram fitting, providing accurate and uniform TRGB distances, and through population synthesis modeling of multi-filter broad- and mediumband photometry.

***Tier 4 – cf. Jansen et al SWP, "Galaxy Assembly and SMBH/AGN Growth from Cosmic Dawn to the End of Reionization"***

***Evolution of the Faint-end Slope of the Dwarf Galaxy Luminosity Function***. The faint-end slope of the galaxy LF is systematically steepening at higher redshifts, reaching a slope $|\alpha|=1.8$–$2.0$ at $z\sim6$. This implies that dwarf galaxies collectively could have produced a sufficient number of ionizing photons to complete the reionization of Hydrogen in the universe by z~6. This critically depends on the escape fraction, $f_{esc}$, of far-UV photons from faint dwarf galaxies. The proposed survey, in particular the UV–blue broad-band filters, could answer this question for statistically meaningful samples per redshift bin.

***Tracing the Reionization History using Lyα-Emitters***. Observations so far have failed to settle the issue of whether the amplitude of the Lyα-emitter LF changes between z = 5.7 and z = 6.5, or as extrapolated from the single detection of a Lyα-emitter at z = 6.96. The proposed medium-band surveys will derive their LF as a function of redshift at z~ > 5.5 over a wide area for large statistical samples and definitively address how the reionization of the IGM progressed over time. Furthermore, the data will allow measuring the ages and clustering properties of Lyα-emitters, and, via the faint-end slope of their LF, their contribution to the budget of ionizing photons.

***Light Profiles of Dwarf Galaxies Around Reionization***. The average radial surface brightness profile derived from stacked, intrinsically similar, z≈6, z≈5, and z≈4 objects imply dynamical ages for these dwarf galaxies of 0.1–0.2Gyr at z≈6–4. These 'dynamical' limits to their ages are comparable to age estimates based on their SEDs, suggesting that the starburst that *finished* the H reionization at z ≃6 may have started by a global onset of Pop II star formation at z ≃6.5–7, or < 200Myr before z≈6. The proposed surveys will yield light profiles, color gradients, and dynamical states of ~ >$10^5$ dwarf galaxies at 0.5<z<7, and provide constraints to their ages from their SEDs.

***The Process of Hierarchical Galaxy Assembly***. The process of galaxy assembly may be directly traced as a function of mass and cosmic environment in the redshift range 0.5<z<5. The HST Deep Fields have outlined how galaxies formed over cosmic time, by measuring the distribution over structure and type as a function of redshift. Sub-galactic units appear to have rapidly merged from z ≃6–8 to grow bigger units to z≈1. Galaxies of all types formed over a wide range of cosmic time, but with a notable transition around z~1.0. Merger products started to settle as galaxies with familiar morphologies, and evolved mostly passively since then. The fine details of this process still elude the HST surveys, because of inadequate spatial sampling and/or depth, and because its FoV is too small to provide sufficient statistics. The proposed imaging through multiple near-UV–near-IR filters and grism(s) would yield robust spectrophotometric redshift estimates ($\sigma_z/(1 + z)$<0.02) for > 5×$10^6$ galaxies with $m_{AB}$< 28–30mag, and allow an analysis of their stellar populations (through population synthesis modeling).

***The Growth of Super-Massive Black Holes***. Through a multi-epoch variability study, the proposed surveys will be able to measure the weak AGN fraction in > $10^5$ field galaxies to $m_{AB}$ < 28–30mag at z < 7.6-7.7 directly, and so robustly constrain how exactly growth of spheroids and SMBHs kept pace with the process of galaxy assembly. The panchromatic imagery and robust spectrophotometric redshifts will allow decomposition of the AGN light from that of the underlying galaxy.

**Technical Overview**

***Executive Summary***

The *Star Formation Camera* (*SFC*) is a dual-channel wide field UV/optical imager covering 19'x15' on the sky, at a plate scale of 18mas per pixel. The design has been chosen to take advantage of the wide, well-corrected field provided by a three-mirror anastigmat (TMA) telescope design to provide the essential combination of a wide field with high angular resolution.

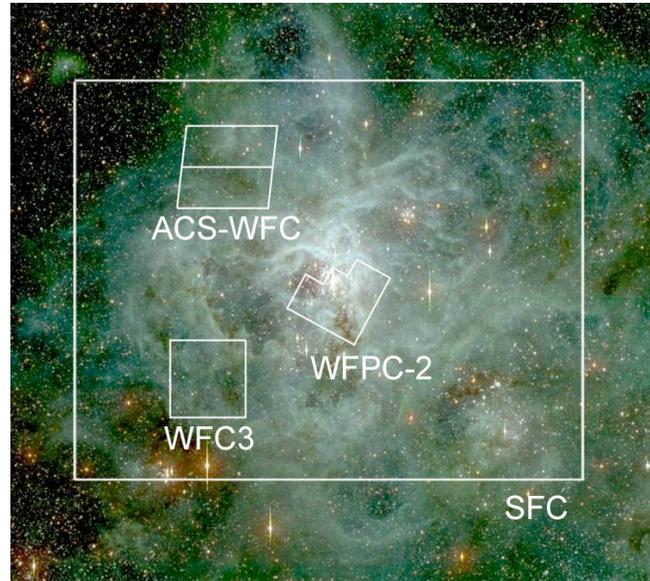

The larger observatory has been designed with a 3 camera fine guidance sensor (FGS) system derived in heritage from the *Spitzer* mission. The *SFC* observing system will use both a star tracker system taking its input from the FGS, and a fast steering mirror (FSM) to achieve the pointing / stability specification of < 0.25 pixel = 4 mas, over the course of 1 exposure (typically ~600 sec.). The camera also possesses an internal calibration channel to allow flat fielding of all filters internal to the camera. The small pixel size will make the sky background very dark and so external flats will be impractical.

The *SFC* science program has been planned and structured to be completed within 1.25 years. Based on our target selection list, our exposure time calculator (ETC) and the likely size of each FPA we have estimated the daily data acquisition rate could be as high as 12 TB, compressible to 6 TB on board. We have projected that downlink transmission of this data load will be attainable by upgrades in DSN well before the targeted launch window of 2021 for such a mission. Using modern and realistic metrics for cost, scale and schedule for the assembly and delivery of the large FPAs that *SFC* will require, we have worked with JPL's Team-X group to provide a 50% confidence cost of $390M FY08 for the development, assembly and flight operation of the camera.

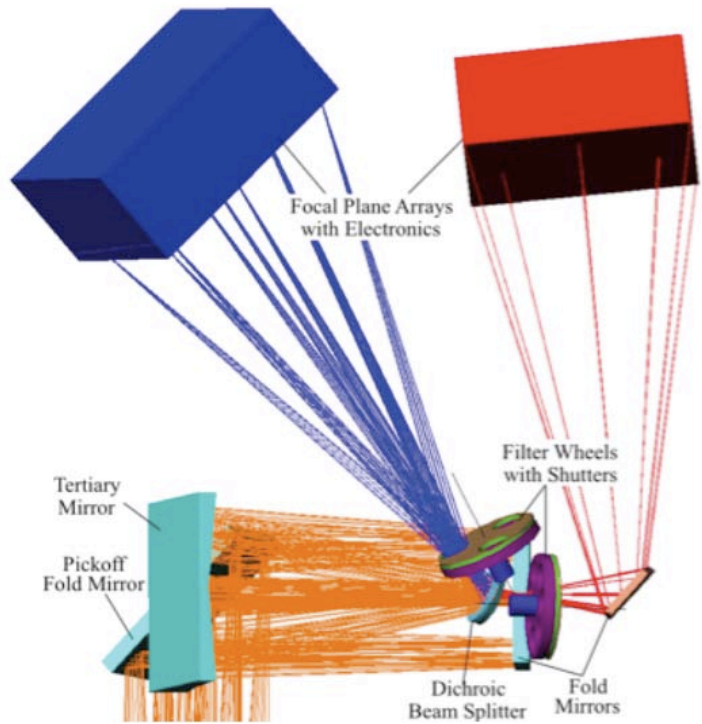

### *Focal Plane Array Design*

The *SFC* employs two large FPAs positioned behind a TMA telescope that provides a large well-corrected field in an off-axis annulus. We have designed *SFC* to use Si CCD chips that have been optimized and custom processed to maximize their response across the full Si-response passband (190-1075nm) – our baseline detectors are the *SNAP*-like LBL 3.5k × 3.5k 10.5μm pixel chips, but post-processed using JPL's Delta Doping technique and AR coated to make each chip as sensitive as possible. We have deliberately chosen to use the same foundry chip for both channels, but with each processed differently to provide optimal response – one type for the Blue (190-517nm) and one type for the Red (517-1075nm). The intent of this decision is to drive down the cost of populating such large FPAs. Having said this it would be entirely possible to build the same FPA using CMOS detectors, but the optical variants of those detectors are not as advanced as the CCDs, at this time. During the next decade this may change.

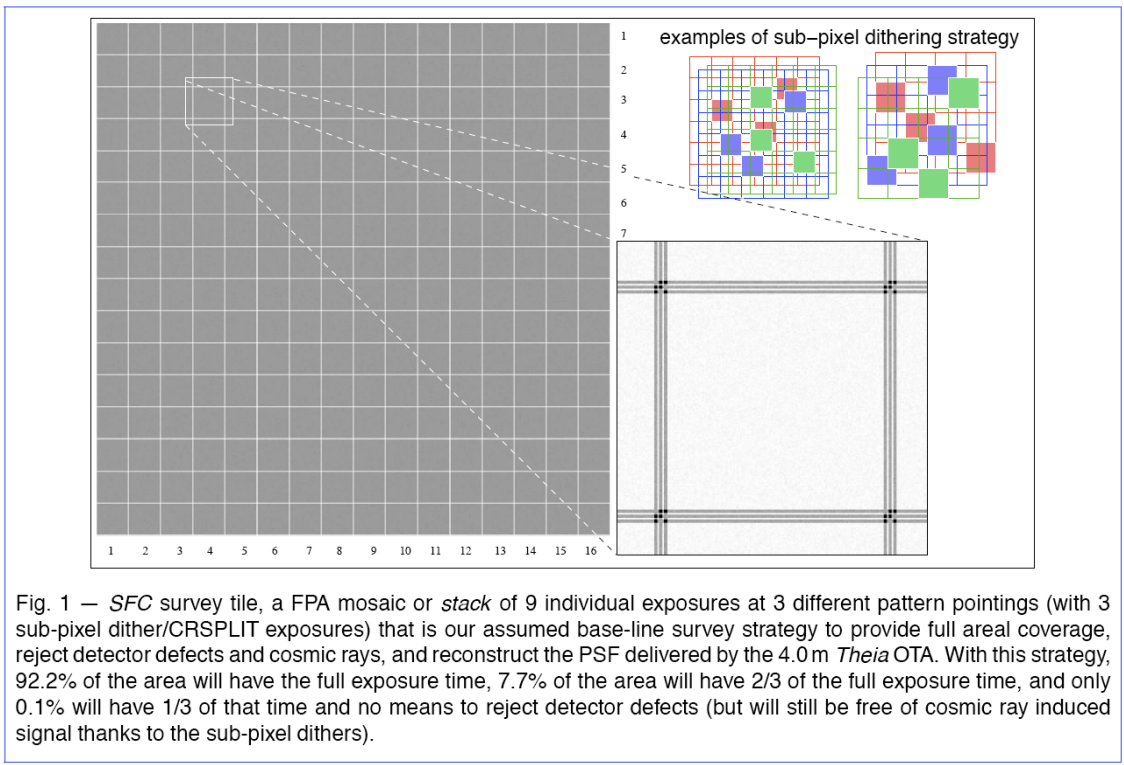

Fig. 1 — *SFC* survey tile, a FPA mosaic or *stack* of 9 individual exposures at 3 different pattern pointings (with 3 sub-pixel dither/CRSPLIT exposures) that is our assumed base-line survey strategy to provide full areal coverage, reject detector defects and cosmic rays, and reconstruct the PSF delivered by the 4.0 m *Theia* OTA. With this strategy, 92.2% of the area will have the full exposure time, 7.7% of the area will have 2/3 of the full exposure time, and only 0.1% will have 1/3 of that time and no means to reject detector defects (but will still be free of cosmic ray induced signal thanks to the sub-pixel dithers).

In designing the optics we have chosen 18 mas pixels, which is the Rayleigh criterion for a 4m mirror at 300nm. The observing strategy we have designed uses a dithering strategy (see Fig 1, achieved by using FSMs) to recover the diffraction limit through Nyquist sampling at the shortest wavelengths, but is also required at longer wavelengths to recover spatial coverage across the interchip seams. This is achieved by requiring each pointing to be comprised of 9 exposures: 3 microdithers at the ¼ pixel level × 3 macrodithers to fill the omitted coverage. This approach will yield a unique combination of a wide field and high spatial resolution to open up parameter space that has heretofor not been available for scientific study.

The current design has a 4:3 aspect delivering a 19'×15' field of view – which corresponds to arrays of 18×15 CCD chips for a total of 270 chips per FPA. *SFC* was designed to provide the right balance between field size and resolution – it is not a perfect solution but it is the most practical – each image taken with *SFC*, with the 2 channels in

parallel – produces twin 6.3 GB images. This means that within a typical 24 hour period, *SFC* could generate as much as 12 TB of data – and as such we have provided for 2 days of on-board data storage by employing lossless compression to reduce the daily data on-disk load to 6 TB. This amount of data can be downlinked in 8-10 hours every 24-hr period using current near-Earth 26 GHz Ka-band technology capacity, and we have specified twice this volume on-board to allow for a potential loss of one downlink opportunity. The CCDs themselves have been specified to operate at a temp of 170-175 K to deliver a dark current of less than 10 e-/pix/hr with a read noise of less than 3 e-, at a gain of 2 e-/ADU with a full-well capacity of 130,000 e-.

***Operational Electronics Design***

In an effort to streamline both the construction of each CCD chip and its support electronics, and to facilitate easier chip swap-out and FPA assembly, the *SFC* focal plane has been designed around the use of a modular stack consisting of the imager, the readout electronics, a FPGA/ASIC controller and associated flash memory. The backplane electronics are dual string for the FPA readout and work with an Electra processor module for control and data readout as well as low noise power supplies for the power supply of the FPAs. The MIC power supply uses a sine wave output, with a fixed square wave chopping DC/DC converter operating at 25 KHz synced to drive the camera readout. Estimated power requirements are 128W per FPA mosaic; with a total electronics power draw of 375W. Electronics mass is estimated at 65Kg.

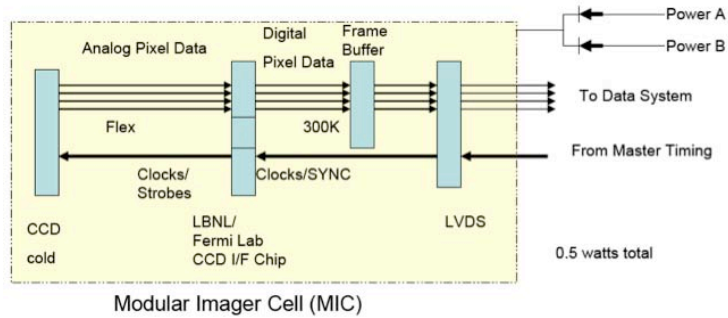

***Thermal Design***

With the goal of assembling so many flight-rated detectors at once, thermal control becomes very challenging. We have specified that the operational temperature of the CCDs needs to be 170-175 K to provide an acceptable dark current rate while mitigating the power necessary to maintain this temperature. To this end variable-conductance heat pipes (VCHP) have been baselined to transport heat from each FPA to the cryogenic radiator. Each FPA has been designed to have one primary and one redundant VCHP. The VCHPs are controllable to reduce conductance for annealing cycles, during which the FPA is held at ~300 K for several hours (to allow removal / mitigation of hot pixel populations).

The Cryoradiator is a structurally integral part of the instrument, mounted with the thermally-isolating structure, and baffled from thermal radiation from the instrument and spacecraft back surface. The radiator is similar in principle to that used on *Kepler*, but larger in size: ~2 m². The radiator occupies ~1/9 of the anti-sun side of spacecraft, and must not view the solar panel back side. The optical bench also has active temperature control to maintain thermal uniformity, for dimensional stability. Each FPA has an associated ASIC package which dissipates 128 W, operating at the temperature of the optical bench (270-300 K). A thermal break is provided by a tape cable 20 cm long, passing through the floating shield. Heat is transported from the ASICs and electronics via heat pipes to the warm radiators on the instrument outer surface.

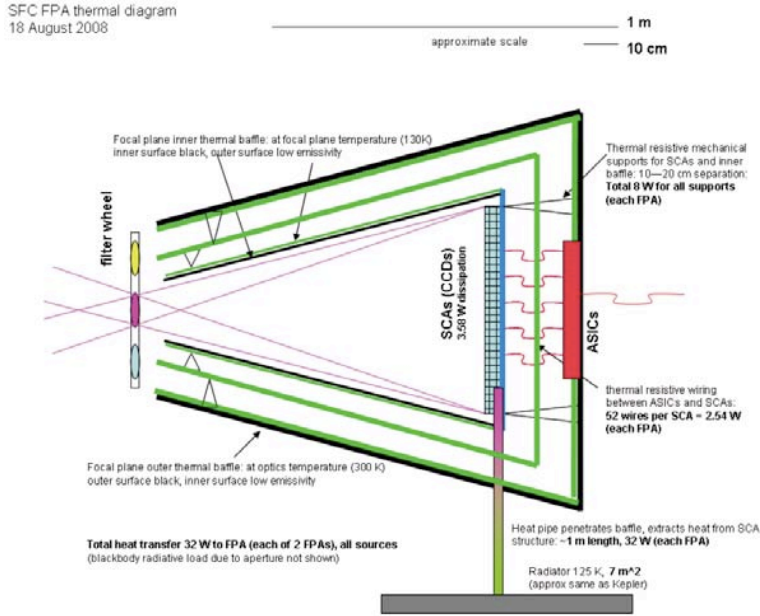

The two FPAs are baffled from thermal radiation from the surrounding optical instrument by a two-level shield: the outer shield is at the instrument optical bench temperature, with black outer and low-emissivity inner surfaces. The inner baffle has low-emissivity outer and black inner surfaces. One floating shield is between them. The thermal baffle has the minimum opening possible at the beam input, to reduce thermal loading into the baffle.

### *Observing with the SFC*

The *SFC* has been deliberately designed to be a powerful survey instrument for observation and measurement of star formation both local and extragalactic. Some of its features have been driven by this intent. The camera has an extensive science-driven complement of both broad-, medium- and narrow-band filters – the details of which are tabulated below. In addition to these filters, we baseline one low- and one medium-resolution grism per channel.

The camera optical design involves 6 optical bounces (driven by design and packaging

| Filter requirements: | wheels must hold at least 16 blue and 18 red filters | | | | (goal: $2 \times 18$ filters) | | | |
|---|---|---|---|---|---|---|---|---|
| **Blue Channel:** | F212M | F262W | F280N | F330N | F372N | F432W | F486N | F502N |
| | *UV1* | *UV2* | MgII | *u* | [O II] | *B* | H$\beta$ | [O III] |
| | 212.8 | 262.3 | 280.9 | 330.2 | 373.5 | 432.7 | 486.1 | 502.3 |
| | 30.0 | 65.0 | 3.5 | 70.0 | 1.5 | 67.5 | 1.6 | 5.0 |
| | F241X | F278XX | F312X | F385X | F467X | F373N | F470N | F487N |
| | *UVX1* | *UVXX* | *UVX2* | *UB* | *g* | [O II] | He II | H$\beta$ |
| | 241.0 | 278.7 | 312.5 | 385.9 | 467.9 | 374.0 | 470.1 | 487.8 |
| | 68.5 | 124.0 | 67.0 | 80.3 | 89.7 | 4.0 | 4.7 | 4.9 |
| **Red Channel:** | F547M | F612W | F632N | F656N | F658N | F775W | F885W | F956N | F990M |
| | *y* | $\mathcal{R}$ | [O I] | H$\alpha$ | [N II] | *i* | *z* | [S III] | $\mathcal{Y}$ |
| | 547.1 | 612.0 | 632.1 | 656.4 | 658.5 | 775.3 | 885.9 | 956.3 | 989.6 |
| | 47.5 | 81.5 | 6.3 | 2.0 | 2.0 | 100.0 | 110.0 | 9.6 | 52.0 |
| | F578X | F659N | F674N | F707X | F867X | F920M | F948M | F980M | F1020M |
| | $\mathcal{V}$ | (H$\alpha$+[N II]) | [S II] | $\mathcal{I}$ | $\mathcal{Z}$ | Ly$\alpha_{z\sim6.6}$ | Ly$\alpha_{z\sim6.8}$ | Ly$\alpha_{z\sim7.1}$ | Ly$\alpha_{z\sim7.4}$ |
| | 579.7 | 659.5 | 674.7 | 707.4 | 870.6 | 920.0 | 948.0 | 979.7 | 1020.6 |
| | 116.8 | 8.7 | 8.1 | 143.5 | 174.5 | 28.1 | 28.0 | 35.7 | 27.9 |

*For each filter, the four rows list the filter name, an alias or feature the filter aims to capture, the central wavelength (in nm) and FWHM in nm. Most narrow-band filters are sufficiently wide (1%) and centered to accomodate relative velocities with respect to the Sun of $-500 \lesssim cz \lesssim +2500$ km/s. Within our own Galaxy and in the Local Group, H$\alpha$ and [N II] emission must be separable, requiring the narrower F656N and F658N, and also the narrower F372N and F486N filters ($-500 \lesssim cz \lesssim +500$ km/s).

issues) and in combination with the large 4m aperture yields a high throughput system that can in a 2000 second exposure reach down to around $29^{th}$ magnitude in the broader filters, and down to surface brightnesses of $10^{-16}$ ergs/cm$^2$/s/arcsec$^2$ through the narrow band filters.

The camera employs a high fidelity dichroic element to allow parallel observations in both the Blue and Red channels – this element also doubles as a FSM for the Blue channel – the Red channel uses its extra fold mirror as the FSM. Observing guidance for *SFC* is provided by the same 3 fast CCDs mounted close to the Cass-like focal plane behind the primary mirror, that are used to drive the FSM to maintain the jitter histogram at less than the ¼ pixel spec.

The very small pixels in *SFC* make it almost impossible to get decent signal from the sky to allow sky-flats – as such we have included an internal CAL channel to allow flatfielding using a lightpipe and diffuser co-mounted in the first filterwheel on each channel. The rotating shutter is comounted in front of each filter assembly and can deliver exposure times ranging from a fraction of a second up to as long as 2000 seconds, and can be used with ND filters.

### Instrument Concept & Summary

*SFC* is a wide-field dichroic camera designed to allow efficient imaging surveys with a 4m space-based UV/optical telescope (e.g., Spergel et al. AWP "*THEIA*: A 4-m Successor to *HST*"). Its parallel Blue (190--517nm) and Red (517--1075nm) channels allow simultaneous imaging of the same 19'×15' FOV in 2 filters. With appropriate dithering, *SFC* delivers diffraction-limited performance down to 300nm. Each channel has its own filterwheel, shutter, FSM, and focus mechanism. *SFC* can drive the telescope as primary instrument, or be used in parallel mode while the telescope is directed by and dwells on a target of another instrument. This will be an effective mode to obtain deep cosmological images.

### Instrument Properties

| | |
|---|---|
| Operating Wavelengths | 190-1075nm (Si sensitivity passband) |
| Observing Channels | Blue (190-517nm); Red (517-1075nm) |
| Broad-band Filters | F241X, F262W, F278XX, F312X, F330W, F385X, F432W, F467X, F578X, F612W, F707X, F775W, F867X, F885W |
| Medium-band Filters | F212M, F547M, F920M, F948M, F980M, F990M, F1020M |
| Narrow-band Filters | F280N, F373N, F486N, F470N, F487N, F502N, F632N, F656N, F658N, F659N, F674N, F956N |
| Grism Filters | G213L, G402M, G745L, G895M |
| Exposure Times | 0.1 up to 2000 seconds |
| Detectors | LBL "SNAP" 3.5k square CCDs |
| Pixel Size | 10.5 μm = 18 mas |
| Field Size | 19' × 15' = 18 × 15 CCDs |
| Dark Noise | < 10 e-/pix/hr |
| Read Noise | < 3 e- |
| Gain | 2 e-/ADU |
| Full Well Capacity | 130,000 e- |
| Operational Temperature | 175 K |
| Pointing Accuracy (w/ FSM) | < ¼ pixel over 2000 seconds |
| Interchip Gap Size | ~35 pixels |
| Mass | 645 Kg |
| Power | 424 W (of which 375W for FPAs) |
| Cost (inc. 30% reserve) | $390M |
| Instrumental Lifetime | 5-yr baseline, 10-yr design |
| Single Field Exposure Image Size | 6.3 GB × 2 channels |
| On-board capacity | 6 TB |
| Typical lossless compression | Factor of 2-3.5 |
| Broadband Sensitivity | $m_V \sim 29$ in 2000 seconds |
| Narrowband Sensitivity | $10^{-16}$ ergs/cm$^2$/s/arcsec$^2$ in 2000 seconds |

**Technology Drivers – cf. Scowen et al TWP, "*Large Focal Plane Arrays for Future Missions*"**

As part of this study we have studied the problem of tiling two large focal planes. The strong scientific case for the use of large focal plane arrays that combine areal coverage with high resolution has been made in a series of 4 white papers to the Decadal Survey (2 by Scowen et al; 2 by Jansen et al). The *SFC* Team-X study has yielded invaluable insight into cost and yield, as well as the likely problems associated with the production and testing of so many CCD chips (~540 flight-rated devices).

As input to the Decadal Survey we are authoring a technology development white paper on the technical issues we have encountered and the expectation for finding a cost effective solution within the next decade. To achieve the capabilities we believe are necessary to make our science goals attainable, we have to surmount serious technological challenges, requiring corresponding investment over the next decade. The issues that we have identified that need particular attention are: how to mass produce large numbers of chips with low read noise, low dark current and high yield from modern lot run manufacture; how to test large numbers of detectors while preserving the fidelity of the product and mitigating the risk associated with fabricating the final array; investment in facilities and opportunities to provide a path to raise the TRL of these emerging technologies; investment in alliances between government labs and academia; development of new packaging designs to minimize interchip gaps when mosaicing large numbers of detectors; critical assessment of what changes in acceptable specifications for flight-rated detectors should look like in the era of mass production; and the expansion of high-capacity data storage, compression and transmission technology.

For the *SFC* part of the study, we have developed an outline to what we believe is necessary to move from the current state of the art to a production line environment to, in this case, tile twin FPAs with as many as 540 separate flight-rated detectors. First, complete the infrastructure for processes and facilities (detectors, readout, packaging), 1 year; second, fabrication and validation of prototype detector modules (SCA) - procure ASIC readout chips, procure detector fabrication run at foundry, process wafers at JPL, packaging: development and fabrication of 10 units, testing and qualification of the module, two years; third, fabrication and validation and demonstration at a ground observatory of prototype 3×3 raft modules (SCA) - procure ASIC readout chips second iterative lot, procure detector fab run, process wafers at JPL, packaging: development and fabrication of 2 units, cold flatness test, shake and bake, radiation, observation including dewar, thermal, miscellaneous additional instrumentation, two years; fourth, fabrication and validation of prototype FPA by assembling two rafts (SCA), 1 year; and fifth, detector balloon demonstration (parallel with above steps 2-4). Overall schedule is four years for first flight with a second flight in the fifth year. Overall ROM cost is $40M FY08.

In light of all this work we believe the low-risk, low-cost, high-fidelity assembly and integration of large focal plane arrays is a vital area of technological development that needs to be invested in over the next decade to enable not only the THEIA mission but a host of other missions and instrumentation that would also benefit from such a capability.

**Activity Organization, Partnerships and Current Status**

This project evolved out of an earlier Origins Science Probe concept study, which led to the *Star Formation Observatory (SFO)* proposal. The team was selected to study the UV/optical camera component of that proposal, and thus *SFC* was born. The specific observatory study was named *Theia* and is comprised of the *XPC* exoplanet characterization instrument, the *UVS* ultra-violet spectrograph and the *SFC*. This project grew out of the NASA Astrophysics Strategic Mission Concept Study (ASMCS) program. The observatory team consists of scientists and engineers at Princeton, STScI, ASU and many other partner universities, NASA centers (JPL, GSFC), and industry partners (Lockheed Martin, Ball Aerospace, and ITT). The original charge to design a single platform to meet all of the science specifications was originally thought to be too challenging, but we are happy to report that we have arrived at a common design that delivers the capabilities necessary to achieve the various projects' core science programs in the baseline time period of 5 years, while also allowing for a 25% GO science program.

**Activity Schedule**

This chart shows the projected schedule for mission development with a launch at the beginning of 2021. The three tall pole technology items associated with *Theia* itself are emphasized. An immediate investment in technology development is essential for this schedule.

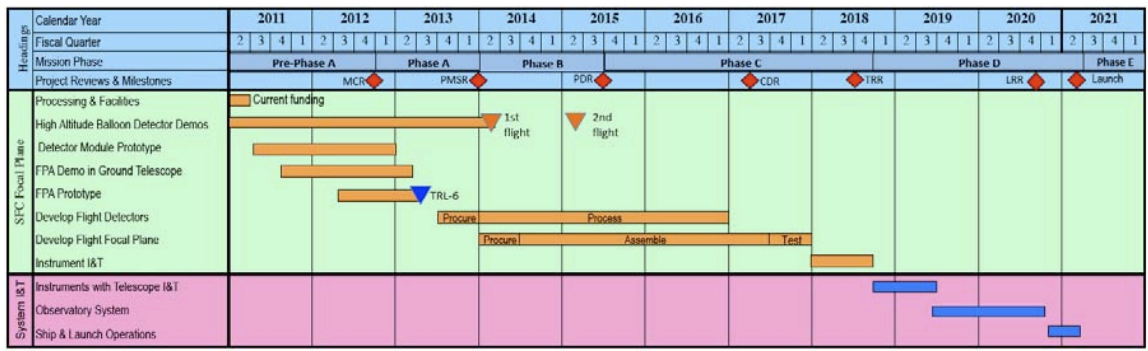

To guarantee the successful development of high capacity production of imaging detectors with the integrity and high yield rate necessary to populate *SFC's* large FPAs, a specific schedule will need to be funded. As can be seen from the timeline above, the start of the *SFC* focal plane development process sets the timeline for the entire observatory.

The development of the capability to produce large FPAs needs to be built up from where the process currently stands. JPL is already working towards large capacity production of chips from wafer sets, and MBE processing of those wafers to perform the delta doping of the semiconductor wafers in sets of as many as 5 at a time. Indeed, the production of the large numbers of chips is really not the challenge – it can be solved by technology alone. The challenge lies in the labor involved in testing and evaluating each of those chips on the optical bench following similar qualification procedures as have been applied to previous flight-rated detectors. Since *SFC* calls for as many as 540 such detectors, this has the potential to become the dominant factor in meeting the delivery deadline.

In the timeline above, we have laid out our best estimate of the realistic timeframes associated with the various phases of development necessary to bring the project to the point of being able to deliver the two *SFC* focal planes. This development starts with current abilities, and moves forward with the testing of many detectors in the lab on the optical bench, then to incorporation of the detectors into a large focal plane ground-based camera, and ultimately with the flight of prototype detectors on an already-proposed balloon-borne telescope. This balloon project has been proposed to lay the scientific ground-work for several aspects of the *SFC* program, but its focal plane will be designed to be similarly modularized to allow flight testing of the detector design, their electronic backplane modules and the command and control of large FPAs at the prototype level. All of these steps require substantial investment and we recommend to the Decadal Survey that such an investment in capability will yield benefits to the community that go far beyond the goals of *SFC*. We are entering an era where the next set of observational breakthroughs will come from the marriage of wide field imaging with high resolution, and it is investment in this kind of development program that will allow such ground-breaking work to be possible.

Even assuming that the FPA prototype can be raised to TRL 6 by mid-2013, while at the same time development via balloon is ongoing, we are still looking at as many as 3-4 years to procure, fabricate, test and assemble the *SFC* FPAs. If we are successful in that regard, we could finish instrument I&T, Phase C and TRR around the end of 2018 allowing a realistic first launch opportunity for such an observatory to be considered during 2021.

This ambitious schedule will require substantial, committed and long-term investment. We believe the benefits to the astronomical community at large would go far beyond the limited boundaries of the project described in this paper, and its associated SWP and TWP documents. We believe that with sufficient vision, astronomers will have new vistas of discovery laid out before them with the utility that such large FPAs could afford.

**Cost Estimates**

As originally instructed by NASA, all participants in the ASMCS program were required to use a common cost estimation model based on the NASA Instrument Cost Model (NICM) by using either the JPL Team-X or GSFC IDC groups to study the proposed projects. *SFC* was studied by JPL's Team-X. The purpose of this mandate was to fulfill the requirement on the part of the Decadal Survey to allow direct comparison between missions so that an educated evaluation of the true cost of any two missions could be made.

The NICM approach uses a database of past missions and instrument costs coupled with a model of current market prices for current technology and services to make a series of costing estimates with varying levels of confidence and contingency. The design requirements laid out in this paper were broken down into various components (mass, power, thermal, mechanical, etc.) and initial estimates for real technological solutions investigated. The results of these studies are laid out in the associated table. In this table the results of this kind of parametric analysis are presented for all aspects of the *Theia* observatory project, but we call attention to the cost estimates for the development of the *SFC* focal plane ($40M FY08), the overall *SFC* instrumentation package (~$390M FY08), and the associated reserves, bus and ATLO expenses for which *SFC* is responsible for a percentage. The Team X cost estimates were generated as part of

| Cost Element | Team-X | THEIA Team | Comments |
|---|---|---|---|
| Pre-Phase A Mission Concept | 15 | 15 | THEIA team estimate |
| Total Technology Development | 150 | 150 | THEIA team estimate |
|    Occulter Starshade | 40 | 40 | |
|    Primary Mirror | 50 | 50 | |
|    SFC Focal Plane | 40 | 40 | |
|    NEXT Ion thruster testing | 20 | 20 | |
| Management, Sys Engr., Mission Design & Mission | 160 | 140 | Derived from other changes |
| Science & Science Data Center | 250 | 250 | |
| Payload System | 3000 | 2200 | |
|    Exoplanet Characterizer | 150 | 150 | |
|    Star Formation Camera | 400 | 400 | |
|    UV Spectrometer | 200 | 200 | Goddard IDC estimate |
|    Fine Guidance Sensors | 80 | 80 | |
|    Occulter Starshield | 100 | 100 | |
|    Other | 25 | 20 | Derived from other changes |
| Telescope | 2000 | 1200 | See text |
| Spacecraft Bus Systems (Observatory & Occulter) | 560 | 500 | THEIA est. uses screened MIL 883B parts (as JWST does), rather than rad-hard Class-S parts |
| System Level I&T (ATLO) | 40 | 100 | THEIA est. adds integration of instruments with telescope & a complex shipping container |
| Mission Operations & Ground Data Systems | 190 | 130 | THEIA est. removes DSN cost, per assumption that THEIA will be an assigned mission |
| Education and Public Outreach | 40 | 35 | Derived from other changes |
| Reserves | 1200 | 1000 | Derived from other changes |
| Launch Services | 440 | 440 | 2 Atlas V 551s |
| Total Mission | **6000** | **5000** | |

a Pre-Phase-A preliminary concept study, are model-based, were prepared without consideration of potential industry participation, and do not constitute an implementation-cost commitment on the part of JPL or Caltech. The accuracy of the cost estimate is commensurate with the level of understanding of the mission concept, typically Pre-Phase A, and should be viewed as indicative rather than predictive.

| Element | Team-X NICM Estimate ($M FY08) |
|---|---|
| Management / Systems Engineering / Product Assurance | 66 |
| I&T | 34 |
| Optics | 21 |
| Electronics | 23 |
| Structures | 16 |
| Thermal | 28 |
| Detectors | 170 |
| Software | 27 |
| Calibration | 5 |
| **Total Instrument** | **390** |

We also include a breakdown of costs for each of the sub-components of the design to illustrate where the majority of the cost comes from – and in short, it is the FPAs.

An aspect of the NICM approach that we did find challenging and that we feel we should call out to the Decadal Survey is that for some aspects of the project design there were substantial inaccuracies introduced by using the formalism strictly. In the case of our large focal planes, the cost estimation to tile the areas needed for *SFC* were estimated by a simple extrapolation of the costs associated with procuring and delivering flight-rated chips for the *WFPC* project in the 1980's. This is still the metric applied to the population of FPAs in new projects, over 20 years later. The true cost of producing the *SFC* focal planes could be very different – partly because the costs and yield rates associated with detector manufacture are so much better now – but also because, as we lay out in the Scowen et al TWP "*Large Focal Plane Arrays for Future Missions*", there is going to have to be some relaxation of selection criteria for literally hundreds of detectors to consider them flight-worthy, otherwise the cost and lead time associated with the delivery of the hardware will become prohibitive.

**Summary**

We have presented an instrument design that provides a technological solution to the specifications laid out in the set of SWPs already filed with the Decadal Survey:

> Scowen et al., SWP, "*Understanding Global Galactic Star Formation*"
> Scowen et al., SWP, "*The Magellanic Clouds Survey – a Bridge to Nearby Galaxies*"
> Jansen et al., SWP, "*A Systematic Study of the Stellar Populations and ISM in Galaxies out to the Virgo Cluster*"
> Jansen et al., SWP, "*Galaxy Assembly and SMBH/AGN Growth from Cosmic Dawn to the End of Reionization*"

The camera takes full advantage of the 4m aperture for which it has been designed to offer world class performance across a truly large focal plane at very high angular resolution. The final design has been studied as part of a concept study and a low-risk, mature solution has been identified that mitigates cost and provides for a realistic schedule to deliver a technologically challenging product.

The challenges associated with the construction of such large FPAs have been discussed within the context of this instrument, but large issues are discussed elsewhere in a TWP already submitted to the Decadal Survey:

> Scowen et al TWP, "*Large Focal Plane Arrays for Future Missions*"

The *SFC* provides a wide-field high-resolution UV/optical dichroic camera that will deliver diffraction-limited images at $\lambda > 300$ nm in both a blue (190-517nm) and a red (517-1075nm) channel simultaneously. It enables and allows the fundamental core science program of conducting a comprehensive and systematic study of the astrophysical processes and environments relevant for the births and life cycles of stars and their planetary systems, and to investigate and understand the range of environments, feedback mechanisms, and other factors that most affect the outcome of the star and planet formation process. The result is an extraordinarily capable instrument that will provide deep, high-resolution imaging across a very wide field enabling an additional wide variety of community science to be pursued.

The technology associated with the camera is next generation but still relatively high TRL, and allowed a low-risk solution to be identified with only moderate technology development investment over the next 10 years.